\documentclass[seceq]{iis}
\usepackage{caption}
\usepackage{color}
\usepackage{ulem}

\newcommand{\ket}[1]{|{#1}\rangle}
\newcommand{\bra}[1]{\langle{#1}|}
\newcommand{\mP}{\mathcal{P}}
\newcommand{\mT}{\mathcal{T}}
\newcommand{\mK}{\mathcal{K}}
\newcommand{\mA}{\mathcal{A}}
\newcommand{\mU}{\mathcal{U}}
\newcommand{\PT}{$\mathcal{PT}$ }

\subjclass{Primary NNXMM, Secondary NNXMM.}

\title{Effects of disorder on non-unitary \PT symmetric quantum walks}
\runtitle{Effects of disorder on non-unitary \PT symmetric quantum walks}

\author{%
\name{Ken \surname{Mochizuki}}
and
\name{Hideaki \surname{Obuse}}
}

\inst{Department of Applied Physics, Hokkaido University, Sapporo
060-8628, Japan}

\runauthor{Ken Mochizuki and Hideaki Obuse}

\abst{\PT symmetry, namely, a combined parity and time-reversal symmetry can make non-unitary quantum walks exhibit entirely real eigenenergy. However, it is known that the concept of \PT symmetry can be generalized and an arbitrary anti-unitary symmetry has a possibility to substitute \PT symmetry. The aim of the present work is to seek such non-unitary quantum walks with generalized \PT symmetry by focusing on effects of spatially random disorder which breaks \PT symmetry. We numerically find non-unitary quantum walks whose quasi-energy is entirely real despite \PT symmetry is broken.}

\kword{\kw{non-unitary quantum walks}, \kw{\PT symmetry}, \kw{randomness}}

\usepackage{amsmath}	
\begin{document}
\maketitle

\section{Introduction}
\label{Introduction}
In the standard quantum mechanics, a Hamiltonian describing a physical system is demanded to possess Hermiticity, which is a sufficient condition to have real eigenenergy. However, non-Hermitian Hamiltonians have been employed to phenomenologically describe open systems in which there are amplification and/or dissipation of particles or energy, namely, gain and/or loss resulting from interactions with the outer environment. In general, the non-Hermitian Hamiltonian has complex eigenenergy. However, in 1998, it was shown that a large number of non-Hermitian Hamiltonians can have entirely real spectra, when they have a combined symmetry of parity symmetry and time-reversal symmetry, that is, \PT symmetry\cite{bender98}. Since this discovery, systems described by non-Hermitian \PT symmetric Hamiltonians have been studied enthusiastically, and for these systems, many peculiar phenomena, such as  unidirectional invisible transport\cite{lin11,mostafazadeh13}, selective single-mode 
 lasings\cite{miri12,feng14}, and so on\cite{hu11,esaki11,bendix09,guo09,longhi09,zheng10,kalish12,garmon15}, which do not appear in Hermitian systems have been theoretically predicted to occur.

In Ref.\ \cite{regensburger12} for the experiment of the quantum walk dynamics realized by coupled optical-fiber loops where effects of gain and loss are highly tunable, the reality of (quasi-)eigenenergy and the unidirectional invisible transport originating from \PT symmetry have been demonstrated. The quantum walk, known as versatile platforms for quantum computations and quantum simulations\cite{kempe03,ambainis03}, is described by time-evolution operators instead of Hamiltonians. Therefore, in the above experiment, the time-evolution operator becomes non-unitary because of effects of gain and loss. Recently, \PT symmetry and the corresponding \PT symmetry operator have been clarified from the non-unitary time-evolution operator directly\cite{mochizuki16}. It is found that in order to retain \PT symmetry, various parameters of the system should satisfy strict conditions not only in space but also in time directions because of parity and time-reversal symmetry, respectivel
 y. However, it is shown that \PT symmetry is one of the sufficient conditions for non-Hermitian Hamiltonians retaining real eigenenergy, and a more general anti-unitary symmetry could restrict eigenenergy of a Hamiltonian to real\cite{bender02,mostafazadeh02}. Thereby, there is a chance to observe  phenomena peculiar to the system with \PT symmetry even in non-unitary quantum walks which do not satisfy the conditions obtained in Ref.\ \cite{mochizuki16}. Especially, if the anti-unitary symmetry is defined as a local operator, unlike \PT symmetry, reality of energy for the system with gain and loss could survive even if spatially random disorder exists in the system.

In the present work, we study effects of spatial disorder for a non-unitary quantum walk which possesses \PT symmetry unless spatial disorder exists. We observe remarkable numerical results that the quasi-eigenenergy remains real even introducing spatial disorder which spoils the condition to establish \PT symmetry. We also consider a non-unitary quantum walk which does not have parity symmetry but retain time-reversal symmetry. We numerically find for this quantum walk  that, while the quasi-eigenenergy for the homogeneous system is complex except a few points, introducing spatial random disorder makes all of  quasi-energy be real. This recovering of reality of quasi-eigenenergy further stimulates work on non-unitary quantum walks.

This paper is organized as follows. In Sec.\ \ref{sec:homogeneous}, we introduce two kinds of non-unitary quantum walks and explain their symmetries. In Sec. \ref{sec:random}, we report our numerical results in non-unitary quantum walks with spatial disorder that the quasi-energy is entirely real in certain parameters, which cannot be predicted from the results in homogeneous non-unitary quantum walks. In Sec.\ \ref{sec:summary}, we summarize the results obtained in this paper.

\section{Non-unitary quantum walks in homogeneous systems and symmetries}
\label{sec:homogeneous}
In this section, we consider homogeneous non-unitary quantum walks and symmetries. At first, in Sec.\ \ref{sec:ptandreality}, we explain the relation between \PT symmetry and reality of the quasi-energy. By generalizing this argument, we show that an arbitrary anti-unitary symmetry could play the same role as \PT symmetry. In Sec.\ \ref{sec:definitionandsymmetries}, we define the time-evolution operator adapted to the experimental setups\cite{regensburger12}. Then, we list conditions and symmetry operators for parity, time-reversal, and \PT symmetries which were obtained in Ref.\ \cite{mochizuki16}. Taking the conditions into account, in Sec.\ \ref{sec:quasi-energy}, we define two kinds of non-unitary quantum walks, \PT symmetric quantum walks and quantum walks with time-reversal symmetry, and we derive dispersion relations of these non-unitary quantum walks analytically.

\subsection{\PT symmetry, reality of the quasi-energy for non-unitary
  time-evolution operators, and generalizations}
\label{sec:ptandreality}
In this section, we explain the relation between reality of the quasi-energy and \PT symmetry for time-evolution operators, and generalize the argument on \PT symmetry to arbitrary anti-unitary symmetries. 

First of all, we introduce quasi-energy $\varepsilon_\lambda$ which is defined from the eigenvalue $\lambda$ of the time-evolution operator;
\begin{align}
U\ket{\psi_{\lambda}}=\lambda\ket{\psi_{\lambda}},\quad
\lambda=e^{-i\varepsilon_{\lambda}},
\label{eq:ue}
\end{align}
where $\ket{\psi_\lambda}$ is the corresponding eigenvector.
According to Ref.\ \cite{mochizuki16}, parity, time-reversal, and
\PT symmetries for a time-evolution operator $U$ are defined as
\begin{subequations}\begin{align}
\mP U \mP^{-1}&=U,
\label{eq:pug}\\
\mT U \mT^{-1}&=U^{-1},
\label{eq:tug}\\
(\mP\mT)U(\mP\mT)^{-1}&=U^{-1},
\label{eq:ptug}
\end{align}\label{eq:ptptug}\end{subequations}
respectively. While the parity symmetry operator $\mP$ is a unitary operator and it flips the sign of position, the time-reversal symmetry operator $\mT$ is an anti-unitary operator, containing the complex conjugation $\mK$, and it reverses the direction of time. Thereby, the \PT symmetry operator is also an anti-unitary operator. (Note that the time-evolution operator $U$ does not include complex conjugation $\mK$ though it is non-unitary.)

If a time-evolution operator has $\mP\mT$ symmetry [Eq.\ (\ref{eq:ptug})] and eigenvectors of the time-evolution operator are also those of the $\mP\mT$ symmetry operator, the quasi-energy $\varepsilon_{\lambda}$ is guaranteed
to be real, that is, the absolute value of the eigenvalue $\lambda$ equals to one; $|\lambda|=1$. The latter condition is expressed by
\begin{align}\label{eq:ptevu}
\mP\mT\ket{\psi_{\lambda}}=e^{i\delta}\ket{\psi_{\lambda}},
\end{align}
where $\delta$ is a real number. However, as already discussed for non-Hermitian Hamiltonians\cite{bender02}, the above argument by using the $\mP\mT$ symmetry operator can be generalized to any anti-unitary symmetry operator
$\mA= \mU \mK$, where $\mU$ is a unitary operator. Therefore, the general condition on reality of quasi-eigenenergy of the non-unitary time-evolution operator is summarized as
\begin{subequations}\begin{align}
\label{eq:arbitraryu1}
\mA\,\, U \mA^{-1}&=U^{-1},\\
\mA \ket{\psi_{\lambda}}&=e^{i\delta}\ket{\psi_{\lambda}}.
\label{eq:arbitraryu2}
\end{align}\end{subequations}

\subsection{Definition of the time-evolution operator and symmetries}
\label{sec:definitionandsymmetries}
We apply the above argument to the time-evolution operator of the one-dimensional (1D) two-step quantum walk introduced in Ref.\ \cite{mochizuki16}.
The time-evolution operator we consider here is written down as
\begin{align}
U&=S\,G(\gamma_{2})\,C(\theta_{2})\,S\,G(\gamma_{1})\,C(\theta_{1}),
\label{eq:upi}
\end{align}
where elemental operators, $i.e.$ the coin operators $C(\theta_{i=1,2})$, the shift operator $S$, and the gain/loss operators $G(\gamma_{i=1,2})$, are defined as
\begin{align}
C(\theta_{i})&=\sum_{n}\ket{n}\bra{n}\otimes\tilde{C}(\theta_{i})=
\sum_{k}\ket{k}\bra{k}\otimes\tilde{C}(\theta_{i}),\,\,\,
\tilde{C}(\theta_{i})=\left(\begin{array}{cc}
\cos\theta_{i}&i\sin\theta_{i}\\
i\sin\theta_{i}&\cos\theta_{i}
\end{array}\right)=e^{i\theta_{i}\sigma_{1}},\label{eq:ch}\\
S&=\sum_{n}\left(\begin{array}{cc}
\ket{n-1}\bra{n}&0\\
0&\ket{n+1}\bra{n}
\end{array}\right)=\sum_{k}\ket{k}\bra{k}\otimes\tilde{S}(k),\,\,\,\
\tilde{S}(k)=\left(\begin{array}{cc}
e^{+ik}&0\\
0&e^{-ik}
\end{array}\right)=e^{ik\sigma_{3}},\label{eq:s}\\
G(\gamma_{\,\,i})&=\sum_{n}\ket{n}\bra{n}\otimes\tilde{G}(\gamma_{\,\,i})=\sum_{k}\ket{k}\bra{k}\otimes\tilde{G}(\gamma_{\,\,i})
,\,\,\,\tilde{G}(\gamma_{\,\,i})=\left(\begin{array}{cc}
e^{\gamma_{\,i}}&0\\
0&e^{-\gamma_{\,i}}
\end{array}\right)
=e^{\gamma_{\,i}\sigma_{3}}.\label{eq:g}
\end{align}
We use the basis of the walker$^{\prime}$s 1D position space $\ket{n}$, momentum space $\ket{k}$, and internal states $\ket{L}=(1,0)^{T}$, $\ket{R}=(0,1)^{T}$ where the superscript T denotes the transpose. The momentum representation can be obtained by applying the Fourier transformation, and $\sigma_{i=1,2,3}$ are the Pauli matrices:
\begin{align*}
\sigma_{1}=\left(\begin{array}{cc}
0&1\\
1&0
\end{array}\right),\,\,\,\,
\sigma_{2}=\left(\begin{array}{cc}
0&-i\\
i&0
\end{array}\right),\,\,\,\,
\sigma_{3}=\left(\begin{array}{cc}
1&0\\
0&-1
\end{array}\right).
\end{align*}
Note that, in the present work, we follow a rule that an operator with a tilde ($\tilde{\,\,}$) on the top acts on space of internal states of walkers. The gain/loss parameters $\gamma_{i}$ take real values, and the gain/loss operators $G(\gamma_{i})$ make the time-evolution operator $U$ non-unitary as long as $\gamma_{i}\neq0$. The gain/loss operators $G(\gamma_{i})$ amplify (attenuate) the wave function amplitudes of left (right) mover components by factors $e^{\gamma_{i}}$ ($e^{-\gamma_{i}}$) when $\gamma_{i}\,>\,0$. 

Here, we briefly explain the argument on symmetries of the non-unitary quantum walk clarified in Ref.\ \cite{mochizuki16}. By using Eqs.\ (\ref{eq:upi})-(\ref{eq:g}), the time-evolution operator in the momentum representation is written as
\begin{align}\label{eq:uk}
U=\sum_{k}\ket{k}\bra{k}\otimes\tilde{U}(k),\,\,\,
\tilde{U}(k)=\tilde{S}(k)\,\tilde{G}(\gamma_{2})\,\tilde{C}(\theta_{2})\,\tilde{S}(k)\,\tilde{G}(\gamma_{1})\,\tilde{C}(\theta_{1}).
\end{align}
In order to make it clear to show symmetries which include inverse of the time-evolution operator in its symmetry relation, we introduce a concept of symmetry time frames\cite{asboth12},
\begin{align}\label{eq:upk}
\tilde{U}^{\prime}(k)=\tilde{C}(\theta_{1}/2)\,\tilde{S}(k)\,\tilde{G}(\gamma_{2})\,\tilde{C}(\theta_{2})\,\tilde{G}(\gamma_{1})\,\tilde{S}(k)\,\tilde{C}(\theta_{1}/2),
\end{align}
which is obtained by the unitary transformation; $\tilde{U}^{\prime}(k)=e^{i\frac{\theta_{1}}{2}\sigma_{1}}\tilde{U}(k)e^{-i\frac{\theta_{1}}{2}\sigma_{1}}$\cite{mochizuki16}. Here, we use the commutative property of $\tilde{S}(k)$ and $\tilde{G}(\gamma_{i})$ as they are described by exponentials of $\sigma_{3}$. Also, separating position or momentum space and space of internal states, we write symmetry operators $\mathcal{P}$, $\mathcal{T}$, and $\mathcal{PT}$ as
\begin{subequations}\begin{align}
\mP&=\sum_{n}\ket{-n}\bra{n}\otimes\tilde{\mP}
=\sum_{k}\ket{-k}\bra{k}\otimes\tilde{\mP},\label{eq:pot}\\
\mT&=\sum_{n}\ket{n}\bra{n}\otimes\tilde{\mT}
=\sum_{k}\ket{-k}\bra{k}\otimes\tilde{\mT},\label{eq:tot}\\
\mP\mT&=\sum_{n}\ket{-n}\bra{n}\otimes\tilde{\mP}\tilde{\mT}
=\sum_{k}\ket{k}\bra{k}\otimes\tilde{\mP}\tilde{\mT}.\label{eq:ptot}
\end{align}\label{eq:symmetry operators}\end{subequations}
Note that, symmetry operators in space of internal states, $\tilde{\mathcal{P}}$, $\tilde{\mathcal{T}}$, and $\tilde{\mathcal{P}}\tilde{\mathcal{T}}$, are independent of position $n$ and momentum $k$, and do not change the sign of $n$ and $k$. From Eqs.\ (\ref{eq:ptptug}) and (\ref{eq:symmetry operators}), we understand that the time-evolution operator in the momentum space needs to satisfy
\begin{subequations}
\begin{align}
\tilde{\mP}\tilde{U}^{\prime}(k)\tilde{\mP}^{-1}&=\tilde{U}^{\prime}(-k),\label{eq:put}\\
\tilde{\mT}\tilde{U}^{\prime}(k)\tilde{\mT}^{-1}&=\tilde{U}^{\prime-1}(-k),\label{eq:tut}\\
(\tilde{\mP}\tilde{\mT})\tilde{U}^{\prime}(k)(\tilde{\mP}\tilde{\mT})^{-1}&=\tilde{U}^{\prime-1}(+k),\label{eq:ptut}
\end{align}\label{eq:symmetry relations}\end{subequations}
to have each symmetry. Here we assume that conditions to retain the above symmetries disassemble into those for each elemental operator in Eqs.\ (\ref{eq:ch})-(\ref{eq:g}). By substituting Eq.\ (\ref{eq:upk}) into Eq.\ (\ref{eq:tut}), for example, the left and right hand sides of Eq. (\ref{eq:tut}) become 
\begin{align}
\text{LHS}&=[\tilde{\mT}\tilde{C}(\theta_{1}/2)\tilde{\mT}^{-1}][\tilde{\mT}\tilde{S}(k)\tilde{\mT}^{-1}][\tilde{\mT}\tilde{G}(\gamma_2)\tilde{\mT}^{-1}][\tilde{\mT}\tilde{C}(\theta_{2})\tilde{\mT}^{-1}]
[\tilde{\mT}\tilde{G}(\gamma_1)\tilde{\mT}^{-1}]
[\tilde{\mT}\tilde{S}(k)\tilde{\mT}^{-1}]
[\tilde{\mT}\tilde{C}(\theta_{1}/2)\tilde{\mT}^{-1}],\nonumber\\
\text{RHS}&=[\tilde{C}^{-1}(\theta_{1}/2)][\tilde{S}^{-1}(-k)][\tilde{G}^{-1}(\gamma_1)][\tilde{C}^{-1}(\theta_{2})]
[\tilde{G}^{-1}(\gamma_2)][\tilde{S}^{-1}(-k)]
[\tilde{C}^{-1}(\theta_{1}/2)],\nonumber
\end{align}
respectively.
By comparing the above two equations, and 
taking account of  relations $\tilde{C}^{-1}(\theta)=\tilde{C}(-\theta)$,
$\tilde{S}^{-1}(-k)=\tilde{S}(+k)$, and
$\tilde{G}^{-1}(\gamma)=\tilde{G}(-\gamma)$ 
acquired from Eqs. (\ref{eq:ch})-(\ref{eq:g}), 
we can obtain sufficient conditions to retain time-reversal symmetry for elemental operators (see Appendix \ref{sec:appentix} for more details). Iterating the same way for the other symmetries, we obtain sufficient conditions for elemental operators $\tilde{C}(\theta_{i})$,
$\tilde{S}(k)$ and $\tilde{G}(\gamma_{\,\,i})$ to retain each symmetry:
\begin{subequations}
\begin{align}
\tilde{\mP}\tilde{C}(\theta_{i})\tilde{\mP}^{-1}&=\tilde{C}(+\theta_{i})&
\tilde{\mP}\tilde{S}(k)\tilde{\mP}^{-1}&=\tilde{S}(-k), &
\tilde{\mP}\tilde{G}(\gamma_{\,i})\tilde{\mP}^{-1}&=\tilde{G}(+\gamma_{\,i}),
\label{eq:pcsgt}\\
\tilde{\mT}\tilde{C}(\theta_{i})\tilde{\mT}^{-1}&=\tilde{C}(-\theta_{i}),&
\tilde{\mT}\tilde{S}(k)\tilde{\mT}^{-1}&=\tilde{S}(+k),
 &
\tilde{\mT}\tilde{G}(\gamma_{\,i})\tilde{\mT}^{-1}&=\tilde{G}(-\gamma_{\,j}),
\label{eq:tcsgt}\\
(\tilde{\mP}\tilde{\mT})\tilde{C}(\theta_{i})(\tilde{\mP}\tilde{\mT})^{-1}&=\tilde{C}(-\theta_{i}),&
(\tilde{\mP}\tilde{\mT})\tilde{S}(k)(\tilde{\mP}\tilde{\mT})^{-1}
&=\tilde{S}(-k),&
(\tilde{\mP}\tilde{\mT})\tilde{G}(\gamma_{\,i})(\tilde{\mP}\tilde{\mT})^{-1}&=\tilde{G}(-\gamma_{\,j}),\label{eq:ptcsgt}
\end{align}\label{eq:elemental symmetry relations}\end{subequations}
where $i,j=1,2$ and $i\neq j$. Note that the relations for the shift and gain/loss operators are determined up to sign, since $\tilde{S}(k)$ and $\tilde{G}(\gamma_i)$ appear twice in the two-step quantum walk in Eq.\ (\ref{eq:upk}). In the case $\gamma_{\,i}=0$ (the unitary quantum walk), the time-evolution operator preserves parity symmetry, time-reversal symmetry, and \PT symmetry, which is confirmed by employing the following symmetry operators:
\begin{subequations}
\begin{align}
\tilde{\mP}&=\sigma_{1},\\
\tilde{\mT}&=\sigma_{1}\mK,
\label{eq:t}\\
\tilde{\mP}\tilde{\mT}&=\sigma_{0}\mK,
\end{align}\label{eq:ptptt}\end{subequations}
where $\sigma_0$ is a $2 \times 2$ identity matrix.

\subsection{Non-unitary quantum walk with \PT symmetry and one with time-reversal symmetry}
\label{sec:quasi-energy}
Now, we focus on contributions of the gain/loss operators, which make the quantum walks non-unitary, by considering the symmetry relations for the gain/loss operators in Eq.\ (\ref{eq:elemental symmetry relations}).

From the symmetry relation for the gain/loss operators $\tilde{G}(\gamma_{\,i})$ in Eq.\ (\ref{eq:ptcsgt}), we find that $\gamma_{1}=-\gamma_{2}$ should be satisfied to retain $\mP\mT$ symmetry. Therefore, the time-evolution operator for the 1D two-step quantum walk with \PT symmetry becomes 
\begin{align}
U_{1}&=S\,G(+\gamma)\,C(\theta_{2})\,S\,G(-\gamma)\,C(\theta_{1}),
\label{eq:u1pi}
\end{align}
where we assume $\gamma > 0$ for simplicity. We note that the above choice of $\gamma_1=-\gamma_2$ does not satisfy the relations for the gain/loss operators in Eqs. (\ref{eq:pcsgt}) and (\ref{eq:tcsgt}). Then, the time-evolution operator $U_{1}$ preserves neither parity symmetry nor time-reversal symmetry.

In addition to the \PT symmetric non-unitary quantum walk, we consider the other time-evolution operator. Since, as explained in Sec.\ \ref{sec:ptandreality}, any anti-unitary symmetry operator may make the quasi-eigenenergy real, it is interesting to consider the non-unitary quantum walk which has time-reversal symmetry defined by the anti-unitary operator [Eq.\ (\ref{eq:t})]  but no parity and \PT symmetries. Taking account of Eq.\ (\ref{eq:tcsgt}), therefore, $\gamma_{1}=\gamma_{2}$ should be satisfied to retain time-reversal symmetry and we compose the non-unitary quantum walk with time-reversal symmetry
\begin{align}
U_{2}&=S\,G(+\gamma)\,C(\theta_{2})\,S\,G(+\gamma)\,C(\theta_{1}).
\label{eq:u2pi}
\end{align}

In the two time-evolution operators, $U_{1}$ and $U_{2}$, effects of gain and loss are included in different ways. On one hand, the time-evolution operator $U_{1}$ attenuates (amplifies) wave function amplitudes of left (right) mover components by factors $e^{-\gamma}$ ($e^{\gamma}$) at the former half of the single time step, and vice verse at the latter half. On the other hand, the time-evolution operator $U_{2}$ keeps amplifying (attenuating) wave function amplitudes of left (right) mover components by factors $e^{\gamma}$ ($e^{-\gamma}$).

\begin{figure}[t]
\centering
\includegraphics[width=7cm]{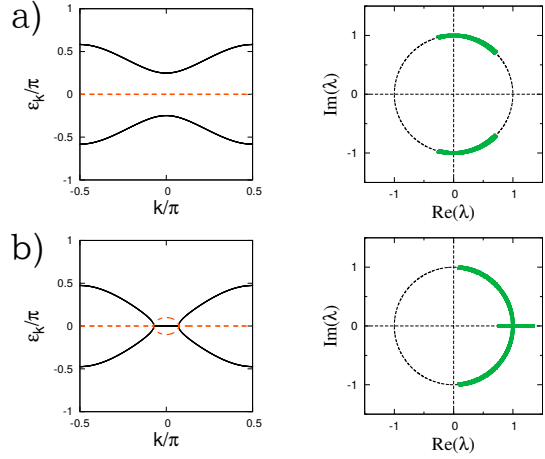}
\caption{The quasi-energy $\varepsilon_{k}$ and eigenvalue $\lambda=e^{-i\varepsilon_{k}}$ of $U_{1}$ with two gain/loss parameters when $\theta_{1}=\pi/3$ and $\theta_{2}=-\pi/12$. The left column shows the quasi-energy $\varepsilon_{k}$ as a function of momentum $k$. The solid (dashed)
curves represent the real (imaginary) part of $\varepsilon_{k}$. The right column shows the eigenvalue $\lambda$ in the complex plane. The circle described by the dashed line represents the unit circle. a) In the case $e^{\gamma}=1.1$, the imaginary part is always zero, and all eigenvalues are on the unit circle. b) In the case $e^{\gamma}=2.2$, part of the quasi-energy becomes imaginary and eigenvalues are not on the unit circle.}
\label{fig:qe1}
\end{figure}

\begin{figure}[t]
\centering
\includegraphics[width=7cm]{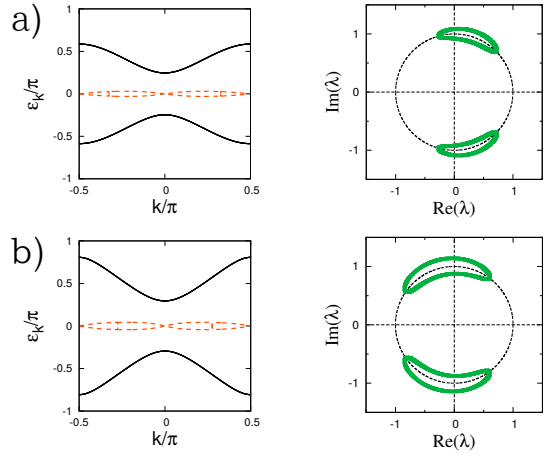}
\caption{The quasi-energy $\varepsilon_{k}$ and eigenvalue $\lambda=e^{-i\varepsilon_{k}}$ of $U_{2}$ with a) $\theta_1=\frac{\pi}{3}$, $\theta_2=-\frac{\pi}{12}$ (the same with Fig. \ref{fig:qe1}) and b) $\theta_1=\frac{\pi}{4}$, $\theta_2=\frac{\pi}{20}$. For both cases, we fix  $e^{\gamma}=1.1$. The left column shows the quasi-energy $\varepsilon_{k}$ as a function of momentum $k$. The solid (dashed) curves represent the real (imaginary) part of $\varepsilon_{k}$. The right column shows the eigenvalue $\lambda$ in the complex plane. The circle described by the dashed line represents the unit circle. In both cases, almost all of quasi-eigenenergies (eigenvalues) are complex (not on the unit circle).}
\label{fig:qe2}
\end{figure}

The quasi-eigenenergy for both time-evolution operators in the homogeneous system is derived by substituting elemental operators in the momentum representation in Eqs.\ (\ref{eq:ch})-(\ref{eq:g}) into Eqs. (\ref{eq:u1pi}) and (\ref{eq:u2pi}). The quasi-energy $\varepsilon_k$ for the time-evolution operator $U_1$ becomes
\begin{align}\label{eq:qe1}
\cos(\pm\varepsilon_k)=\cos\theta_{1}\cos\theta_{2}\cos2k
-\sin\theta_{1}\sin\theta_{2}\cosh(2\gamma),
\end{align}
which is shown in Fig. \ref{fig:qe1} with certain parameters. We understand that when the absolute value of the right hand side in Eq.\ (\ref{eq:qe1}) is smaller than or equal to one the quasi-energy $\varepsilon_k$ remains real. Beyond a certain value of $\gamma$, this condition is broken and then part of quasi-energy becomes complex as shown in Fig.\ \ref{fig:qe1}.
In the case of the time-reversal symmetric quantum walk $U_2$, the quasi-energy becomes
\begin{align}\label{eq:qe2}
\cos(\pm\varepsilon_k)=\cos\theta_{1}\cos\theta_{2}\cosh(2\gamma)\cos2k
-\sin\theta_{1}\sin\theta_{2}+i\cos\theta_{1}\cos\theta_{2}\sinh(2\gamma)\sin2k.
\end{align}
In contrast to Eq.\ (\ref{eq:qe1}), there is the imaginary term in the right hand side of Eq.\ (\ref{eq:qe2}). Thereby, if the gain/loss parameter $\gamma$ takes any finite value except at $\theta_1,\theta_2=\pm\frac{\pi}{2}$, $\varepsilon_k$ becomes complex, which is shown in Fig.\ \ref{fig:qe2}. This indicates that the non-unitary quantum walk $U_2$ does not satisfy the condition Eq.\ (\ref{eq:arbitraryu2}) with $\mA=\mT$. This seems to suggest that the time-evolution operator $U_2$ is unsuitable to study reality of quasi-energy of non-unitary quantum walks. However, this is not the case, as we demonstrate below.

\section{Non-unitary quantum walks with spatially random coin operators}
\label{sec:random}
\begin{figure}[htbp]
\centering
\includegraphics[width=8cm]{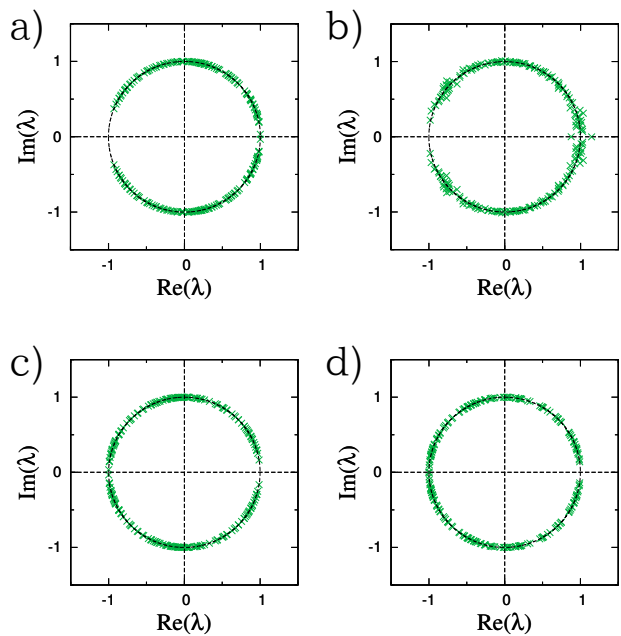}
\caption{The eigenvalues $\lambda=e^{-i\varepsilon}$ of the
 time-evolution operators $U_{1}$ and $U_{2}$ in (a) case-A, (b) case-B, (c) case-C, and (d) case-D in the complex plane. The mean values $\overline{\theta}_1$ and $\overline{\theta}_2$ ($\theta_2$) are the same with $\theta_1$ and $\theta_2$ in the homogeneous cases written in captions in Fig.\ \ref{fig:qe1} and Fig. \ref{fig:qe2} b), and the gain/loss parameter is $e^{\gamma}=1.1$\cite{explain}. The circle described by the dashed curve represents the
 unit circle.}
\label{fig:caseabcd}
\end{figure}

Finally, we study effects of disorder on the time-evolution operators $U_1$ and $U_2$. In this work, we assume that disorder comes only from spatial fluctuations of the angle $\theta$ of the coin operator, thus, we modify the coin operator in Eq.\ (\ref{eq:ch}) by allowing the angle to depend on position,
\begin{align}
C(\theta_{i})&=\sum_{n}\ket{n}\bra{n}\otimes\tilde{C}[\theta_{i}(n)],\quad
\tilde{C}[\theta_{i}(n)]=\left(\begin{array}{rr}
\cos\theta_{i}(n)&i\sin\theta_{i}(n)\\ 
i\sin\theta_{i}(n)&\cos\theta_{i}(n)
\end{array}\right).
\label{eq:cr}
\end{align}
At first, we confirm whether the random angle $\theta_i(n)$ affects symmetry of the time-evolution operators $U_1$ and $U_2$. By the same way as we verified symmetries of the time-evolution operator in the homogeneous system, we derive  symmetry relations for the position dependent coin operator to establish \PT symmetry and  time-reversal symmetry as $(\mP\mT) C(\theta_i)  (\mP \mT)^{-1} = C(-\theta_i)$ and $\mT C(\theta_i) \mT^{-1} = C(-\theta_i)$, respectively, with the same symmetry operators in Eqs.\ (\ref{eq:symmetry operators}) and (\ref{eq:ptptt}). Due to the parity symmetry operator, the condition for the angle $\theta_i(n)$ to retain \PT symmetry is
\begin{align}
\theta_{i}(n)=\theta_{i}(-n),\label{eq:ptct}
\end{align}
while the one to retain time-reversal symmetry is trivial
$[\theta_{i}(n)=\theta_{i}(n)]$ since the time-reversal symmetry operator $\mT$ is the local operator. This gives distinct consequences on symmetries of the time-evolution operators $U_1$ and $U_2$ if disorder exists. In the case of $U_2$, time-reversal symmetry of the time-evolution operator $U_{2}$ is preserved even when  $\theta_i(n)$ is uncorrelated random angles in position space. However, in the case of $U_1$, the uncorrelated random angle is
inconsistent with Eq.\ (\ref{eq:ptct}), and then \PT symmetry is broken.
\begin{table*}[htbp]
\begin{center}
\caption{Four cases, A-D, of non-unitary quantum walks with spatially disordered angles of the coin operator studied in Sec. \ref{sec:random}. This table should be read, $i.e.$ in case-A, the time-evolution operator $U_1$ in
 Eq. (\ref{eq:u1pi}) is assigned to contain the disordered $\theta_{1}$ and constant $\theta_{2}$.}
\scalebox{1.2}{
\begin{tabular}{cccc}
\hline\hline
case & time-evolution operator & $\theta_{1}$ & $\theta_{2}$  \\ \hline
 A & $U_1$ & random & constant  \\ 
 B & $U_1$ & random & random  \\ 
 C & $U_2$ & random & constant  \\ 
 D & $U_2$ & random & random  \\ \hline\hline
\label{tb:cases}
\end{tabular}
}
\end{center}
\end{table*}
Nevertheless, we consider effects of spatially disordered $\theta_i(n)$ on both
time-evolution operators $U_1$ and $U_2$. We treat the random $\theta_i(n)$ obeying the box distribution as follows;
\begin{equation}
\theta_i(n) \in [\overline{\theta}_{i}-\pi/4,\,\, \overline{\theta}_{i}+\pi/4], 
\end{equation}
where $\overline{\theta}_{i}$ is the mean value of the distributed $\theta_{i}(n)$. Since there are two coin operators in the time-evolution operators $U_1$ and $U_2$, we consider four cases as listed in Table \ref{tb:cases}. We numerically calculate eigenvalues of the non-unitary quantum walks $U_1$ and
$U_2$ in the finite position space ($120$ nodes) by imposing periodic boundary
conditions on both ends of the system. Figure \ref{fig:caseabcd} shows eigenvalues of cases A-D for a single disorder realization at certain parameters. In case-A where only $\theta_1$ is random in $U_1$ [Fig. \ref{fig:caseabcd} (a)], all eigenvalues of $U_{1}$ are on the unit circle in the complex plane (quasi-eigenenergy is entirely real) even though the time-evolution operator $U_{1}$ does not preserve \PT symmetry. This suggests that there should be more generalized \PT symmetry for the time-evolution operator in case-A which we have not yet identified. However, we remark that, in case-B [Fig. \ref{fig:caseabcd} (b)] where both $\theta_1$ and $\theta_2$ are random, all eigenvalues deviate from the unit circle. Therefore, the quasi-eigenenergy becomes entirely complex.

Contrary, in case-C [Fig. \ref{fig:caseabcd} (c)] and case-D [Fig. \ref{fig:caseabcd} (d)], the quasi-eigenenergy of $U_{2}$ with spatially disordered coin operators becomes entirely real. Recalling that almost all quasi-eigenenergy are complex when both $\theta_{1}$ and $\theta_{2}$ are constant (see Fig.\ \ref{fig:qe2}), this indicates recovering of reality of quasi-eigenenergy induced by random angles $\theta$. Moreover, in case-C and case-D, we numerically confirm that all eigenvectors of $U_{2}$ are those of time-reversal symmetry operator $\mT$ in Eqs. (\ref{eq:tot}) and (\ref{eq:t}), that is, reality of the quasi-eigenenergy results from time-reversal symmetry, satisfying Eq. (\ref{eq:arbitraryu2}) with $\mA=\mT$. 

\begin{figure}[htbp]
\centering
\includegraphics[width=15.5cm]{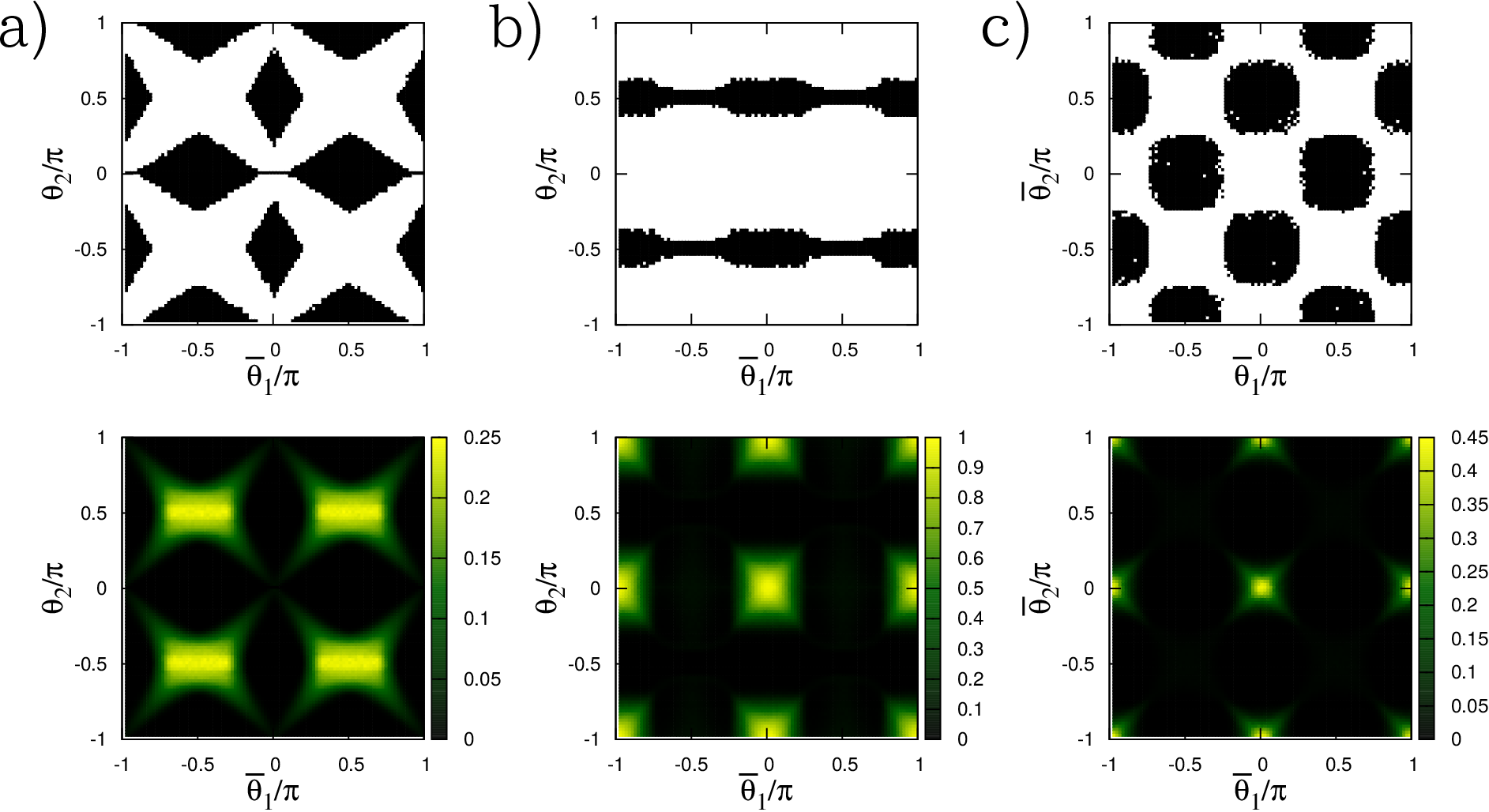}
\caption{(top) The presence (white) or absence (black) of complex quasi-eigenenergy in all eigenstates and (bottom) the ratio of the number of eigenstates with complex quasi-eigenenergy to the number of all eigenstates for the (a) case-A, (b) case-C, and (c) case-D in the parameter space of $\overline{\theta}_1$ and $\theta_2$ (or $\overline{\theta}_2$).
The gain/loss parameter is set to $e^{\gamma}=1.1$. }
\label{fig:colormap}
\end{figure}

These remarkable results are confirmed in wide parameter regions. We show the presence or absence of complex quasi-eigenenergy in all eigenstates [Fig.\ \ref{fig:colormap} (top)] and the ratio of the number of eigenstates with complex quasi-eigenenergy to the number of all eigenstates [Fig.\ \ref{fig:colormap} (bottom)] for the cases A, C, and D for various values of $\overline{\theta}_1$ and $\theta_2$ (or $\overline{\theta}_2$). The number of disorder realizations is 200 for each case. (Note that, since  there is no eigenstate with real quasi-eigenenergy in the case-B, there is no corresponding figure.) We clearly find that entirely real quasi-eigenenergy remains in finite parameter regions for the cases A,C, and D. Note that, in the case C, although the angles $\overline{\theta}_1$ and $\theta_2$ chosen to calculate eigenvalues of a single disorder realization in Fig.\ \ref{fig:caseabcd} (c) are in a white region in Fig.\ \ref{fig:colormap} (b) (top), that is, complex quasi-eigenenergies exist among 200 ensembles, the probability that the time-evolution operator $U_{2}$ has complex quasi-eigenenergy is very small [see Fig.\ \ref{fig:colormap} (b) (bottom)]. Although these are numerical results for the finite system with the finite number of ensembles, we believe that the ratio [Fig.\ \ref{fig:colormap} (bottom)] does not change drastically as increasing the system size and number of disorder realizations.

\section{Summary}
\label{sec:summary}

We have considered the non-unitary quantum walks with spatially disordered coin operators and studied how their quasi-energy is affected. To this end, we have introduced two kinds of non-unitary quantum walks; one has \PT symmetry and the other one has time-reversal symmetry as a possible choice of generalized \PT symmetry. We have shown that the non-unitary quantum walk with \PT symmetry exhibits entirely real quasi-energy in a certain parameter space for the homogeneous system. Remarkably, although the spatially disordered coin operator breaks \PT symmetry, the quasi-energy remains real even if one of the two coin operators in the time-evolution operator in Eq.\ (\ref{eq:u1pi}) is replaced with the disordered one. Furthermore, we have observed  recovering of reality of quasi-energy of the time-reversal symmetric non-unitary quantum walk in Eq.\ (\ref{eq:u2pi}) induced by the spatially disordered coin operator(s). Since there remain several open questions such as identifying
  the generalized \PT symmetry for the time-evolution operator $U_1$ with the disordered coin operator and the reason why  the time-evolution operator $U_2$ recovers reality of quasi-energy by introducing the disordered coin operator, the result in the present work stimulates one to investigate more general symmetry or conditions for non-unitary quantum walks to retain real quasi-energy. Furthermore, since the non-unitary quantum walk can be realized in experiment\cite{regensburger12}, the quantum walk would provide an intriguing arena to study phenomena peculiar to non-unitary systems.

\section*{Acknowledgments}

We thank Y. Asano and K. Yakubo for helpful discussions. This work was supported by the ``Topological Materials
Science'' (No.\ JP16H00975) and Grants-in-Aid (No.\ JP16K17760 and No.\ JP16K05466) from
the Japan Society for Promotion of Science.

\appendix
\setcounter{equation}{0}
\setcounter{table}{0}
\setcounter{figure}{0}
\renewcommand{\theequation}{S\arabic{equation}}
\renewcommand{\thefigure}{S\arabic{figure}}
\renewcommand{\thetable}{S\arabic{table}}
\section{Details of the Derivation of Eq.\ (\ref{eq:tcsgt})}
\label{sec:appentix}
In Appendix \ref{sec:appentix}, we explain details of the deriviation of Eq.\ (\ref{eq:tcsgt}). As written in the main text, the time-evolution operator $U'$ and the time-reversal symmetry operator $\mT$ are described as
\begin{equation}
U'=\sum_k\ket{k}\bra{k}\otimes\tilde{U}'(k)
\label{eq:up}
\end{equation}
where
\begin{equation}
\tilde{U}'(k)=\tilde{C}(\theta_1/2)\tilde{S}(k)\tilde{G}(\gamma_2)\tilde{C}(\theta_2)\tilde{G}(\gamma_1)\tilde{S}(k)\tilde{C}(\theta_1/2),
\label{eq:upks}
\end{equation}
and
\begin{equation}
\mT=\sum_k\ket{-k}\bra{k}\otimes\tilde{\mT},
\label{eq:trso}
\end{equation}
respectively. Note that, $\tilde{\mT}$ is independent of momentum $k$. Also, in order to retain time-reversal symmetry, the time-evolution operator needs to satisfy
\begin{equation}
\mT U' \mT^{-1}=U'^{-1}.
\label{eq:trsr1}
\end{equation}
From Eqs.\ (\ref{eq:up}) and (\ref{eq:trso}), we can understand that the left hand side (LHS) of Eq.\ (\ref{eq:trsr1}) becomes
\begin{eqnarray}
\mT U' \mT^{-1}&=&\sum_k\ket{-k}\bra{-k}\otimes\tilde{\mT}\tilde{U}'(k)\tilde{\mT}^{-1}
\nonumber\\
&=&\sum_k\ket{k}\bra{k}\otimes\tilde{\mT}\tilde{U}'(-k)\tilde{\mT}^{-1}.
\label{eq:LHS1}
\end{eqnarray}
Since the right hand side (RHS) of Eq.\ (\ref{eq:trsr1}) is $U'^{-1}=\displaystyle\sum_k\ket{k}\bra{k}\otimes\tilde{U}'^{-1}(k)$, $\tilde{U}^\prime(k)$ needs to satisfy
\begin{eqnarray}
\tilde{\mT}\tilde{U}'(k)\tilde{\mT}^{-1}=\tilde{U}'^{-1}(-k).
\label{eq:trsr2}
\end{eqnarray}
Substituting Eq.\ (\ref{eq:upks}) into Eq. (\ref{eq:trsr2}), the LHS and RHS of Eq.\ (\ref{eq:trsr2}) are possibly written as 
\begin{align}
\text{LHS}=[\tilde{\mT}\tilde{C}(\theta_{1}/2)\tilde{\mT}^{-1}][\tilde{\mT}\tilde{S}(k)\tilde{\mT}^{-1}][\tilde{\mT}\tilde{G}(\gamma_2)\tilde{\mT}^{-1}][\tilde{\mT}\tilde{C}(\theta_{2})\tilde{\mT}^{-1}]
[\tilde{\mT}\tilde{G}(\gamma_1)\tilde{\mT}^{-1}]
[\tilde{\mT}\tilde{S}(k)\tilde{\mT}^{-1}]
[\tilde{\mT}\tilde{C}(\theta_{1}/2)\tilde{\mT}^{-1}],
\label{eq:LHS2}
\end{align}
and
\begin{align}
\text{RHS}&=[\tilde{C}^{-1}(\theta_{1}/2)][\tilde{S}^{-1}(-k)][\tilde{G}^{-1}(\gamma_1)][\tilde{C}^{-1}(\theta_{2})]
[\tilde{G}^{-1}(\gamma_2)][\tilde{S}^{-1}(-k)]
[\tilde{C}^{-1}(\theta_{1}/2)]\nonumber\\
&=[\tilde{C}(-\theta_{1}/2)][\tilde{S}(+k)][\tilde{G}(-\gamma_1)][\tilde{C}(-\theta_{2})]
[\tilde{G}(-\gamma_2)][\tilde{S}(+k)]
[\tilde{C}(-\theta_{1}/2)],
\label{eq:RHS}
\end{align}
respectively. Then, comparing Eq.\ (\ref{eq:LHS2}) with Eq.\ (\ref{eq:RHS}), we can obtain Eq.\ (\ref{eq:tcsgt}) in the main text:
\begin{align}
\tilde{\mT}\tilde{C}(\theta_i)\tilde{\mT}^{-1}=\tilde{C}(-\theta_i),\ \ \tilde{\mT}\tilde{S}(k)\tilde{\mT}^{-1}=\tilde{S}(k),\ \ 
\tilde{\mT}\tilde{G}(\gamma_i)\tilde{\mT}^{-1}=\tilde{G}(-\gamma_j),\ \ 
\label{eq:sc}
\end{align}
where $i\ne j$, as sufficient conditions to satisfy Eq.\ (\ref{eq:trsr1}), that is, preserve time-reversal symmetry. Also, defining the time-reversal symmetry operator 
\begin{align}
\tilde{\mT}=\sigma_1\mK
\label{eq:to}
\end{align}
as in the main text, we can concretely confirm
\begin{align}
\tilde{\mT}\tilde{C}(\theta_i)\tilde{\mT}^{-1}\tilde{\psi}=\tilde{C}(-\theta_i)\tilde{\psi},\ \ 
\tilde{\mT}\tilde{S}(k)\tilde{\mT}^{-1}\tilde{\psi}=\tilde{S}(k)\tilde{\psi}
\label{eq:tt}
\end{align}
for any vector in coin space $\tilde{\psi}=(\psi_L,\psi_R)^{T}$ in unitary case ($\gamma_1=\gamma_2=0$). 


\begin{thebibliography}{99}


\bibitem{bender98}Bender, C. M. and Boettcher, S., ``Real spectra in non-Hermitian Hamiltonians having PT symmetry'',{\it Phys. Rev. Lett.} {\bf 80}, 5243 (1998).

\bibitem{lin11}Ramezani, Z. Lin, H., Eichelkraut, T., Kottos, T., Cao, H., and Christodoulides, D. N., ``Unidirectional Invisibility Induced by PT-Symmetric Periodic Structures'',{\it Phys. Rev. Lett.} {\bf 106}, 213901 (2011).

\bibitem{mostafazadeh13}Mostafazadeh, A., ``Invisibility and PT symmetry'',{\it Phys. Rev. A.} {\bf 87}, 012103 (2013).

\bibitem{miri12}Miri, M. A., LiKamWa, P., and Christodoulides, D. N., ``Large area single-mode parity-time-symmetric laser amplifiers,{\it Opt. Lett.} {\bf 37}, 764 (2012).

\bibitem{feng14}Feng, L., Wong, Z. J., Wang, X. R-M. Ma, Zhang, Y., ``Single-mode laser by parity-time symmetry breaking'',{\it Science} {\bf 346}, 972 (2014).

\bibitem{hu11}Hu, Y. C., Hughes, T. L., ``Absence of topological insulator phases in non-Hermitian PT-symmetric Hamiltonians'',{\it Phys. Rev. B} {\bf 84}, 153101 (2011).

\bibitem{esaki11}Esaki, K., Sato, M., Hasebe, K., Kohmoto, M., ``Edge states and topological phases in non-Hermitian systems'',{\it Phys. Rev. B} {\bf 84}, 205128 (2011).

\bibitem{bendix09}Bendix, O., R. Fleischmann, Kottos, T., and Shapiro, B., ``Exponentially Fragile PT Symmetry in Lattices with Localized Eigenmodes'',{\it Phys. Rev. Lett.} {\bf 103}, 030402 (2009).

\bibitem{guo09}Guo, A. and Salamo, G. J., ``Observation of PT-Symmetry Breaking in Complex Optical Potentials'',{\it Phys. Rev. Lett.} {\bf 103}, 093902 (2009).

\bibitem{longhi09}Longhi, S., ``Bloch Oscillations in Complex Crystals with PT Symmetry'',{\it Phys. Rev. Lett.} {\bf 103}, 123601 (2009).

\bibitem{zheng10}Zheng, M. C., Christodoulides, D. N., Fleischmann, R. and Kottos, T., ``PT optical lattices and universality in beam dynamics'',{\it Phys. Rev. A} {\bf 82}, 010103 (2010).

\bibitem{kalish12}Kalish, S., Lin, Z., and Kottos, T., ``Light transport in random media with PT symmetry'',{\it Phys. Rev. A} {\bf 85}, 055802 (2012).

\bibitem{garmon15}Garmon, S., Gianfreda, M., Hatano, N., ``Bound states, scattering states, and resonant states in PT-symmetric open quantum systems'',{\it Phys. Rev. A} {\bf 92}, 022125 (2015).

\bibitem{regensburger12}Regensburger, A., Bersch, C., Miri,  M-A., Onishchukov, G., Christodoulides, D. N., and Peschel, U., ``Parity$-$time synthetic photonic lattices'',{\it Nature} {\bf 488} 167 (2012).

\bibitem{kempe03}Kempe, J., ``Quantum random walks: an introductory overview'',{\it Contemp.\  Phys.}\ {\bf 44}, 307 (2003).

\bibitem{ambainis03}
Ambainis,\ A., ``Quantum walks and their algorithmic applications'',{\it Int.\ J.\ Quantum Inform.}\ {\bf 01}, 507 (2003).

\bibitem{mochizuki16}Mochizuki, K.,\ Kim, D., and\ Obuse, H., ``Explicit definition of \PT symmetry for non-unitary quantum walks with gain and loss'', {\it Phys. Rev. A.}\ {\bf 93}, 062116 (2016).

\bibitem{bender02}
Bender, C.\ M.,\ Berry, M.\ V., and\ Mandilara, A., ``Generalized PT symmetry and real spectra'',{\it J.\ Phys.\ A:Math.\ Gen.}\ {\bf 35}, L467 (2002).

\bibitem{mostafazadeh02}
Mostafazadeh, A., ``Pseudo-Hermiticity versus PT Symmetry: The necessary condition for the reality of the spectrum of a non-Hermitian Hamiltonian'',{\it J.\ Math.\ Phys.}\ {\bf 43}, 205 (2002), ``Pseudo-Hermiticity versus PT-Symmetry II: A complete characterization of non-Hermitian Hamiltonians with a real spectrum'',{\it J.\ Math.\ Phys.}\ {\bf 43}, 2814 (2002), `` Pseudo-Hermiticity versus PT-Symmetry III: Equivalence of pseudo-Hermiticity and the presence of antilinear symmetries'',{\it J.\ Math.\ Phys.}\ {\bf 43}, 3944 (2002). 

\bibitem{asboth12}
Asb\'oth, J. K. and Obuse, H., ``Bulk-boundary correspondence for chiral symmetric quantum walks'',{\it Phys. Rev. B.} {\bf 88}, 121406(R) (2013).

\bibitem{explain}
We choose different parameters for $U_1$ and $U_2$, $i.e.$ $\overline{\theta}_1=\frac{\pi}{3}$ and $\overline{\theta}_2(\theta_2)=-\frac{\pi}{12}$ for $U_1$ but $\overline{\theta}_1=\frac{\pi}{4}$ and $\overline{\theta}_2(\theta_2)=\frac{\pi}{20}$ for $U_2$, since we prefer to show the numerical result that eigenvalue of $U_1$ apparently deviate from the unit circle in the complex plane when both $\theta_1$ and $\theta_2$ are randomized. However, we emphasize that our conclusion is unchanged even for other values of $\overline{\theta}_1$ and $\overline{\theta}_2(\theta_2)$ which are shown in Fig. \ref{fig:colormap}.

\end{thebibliography}
\end{document}